%% file: main.tex
\documentclass[journal=nalefd,manuscript=article]{achemso}
\usepackage{chemformula} 
\usepackage[T1]{fontenc} 
\usepackage{braket}
\usepackage{amsmath,amssymb}
\usepackage{tabularx}
\newcolumntype{L}[1]{>{\raggedright\arraybackslash}p{#1}}
\newcolumntype{C}[1]{>{\centering\arraybackslash}p{#1}}
\newcolumntype{R}[1]{>{\raggedleft\arraybackslash}p{#1}}

\newcommand{\hc}{\mathrm{h.c.}}
\usepackage{pdfpages}

\author{Kuniyuki Miwa}
\email{kuniyukimiwa@ims.ac.jp}
\affiliation{Institute for Molecular Science, National Institutes of Natural Sciences, Okazaki 444-8585, Japan}
\affiliation{School of Physical Sciences, Graduate University for Advanced Studies, Okazaki 444-8585, Japan}
\author{Souichi Sakamoto}
\affiliation{Institute for Molecular Science, National Institutes of Natural Sciences, Okazaki 444-8585, Japan}
\author{Akihito Ishizaki}
\email{ishizaki@ims.ac.jp}
\affiliation{Institute for Molecular Science, National Institutes of Natural Sciences, Okazaki 444-8585, Japan}
\affiliation{School of Physical Sciences, Graduate University for Advanced Studies, Okazaki 444-8585, Japan}

\title{Control and enhancement of single-molecule electroluminescence through strong light-matter coupling}

\begin{document}

\begin{abstract}
The energetic positions of molecular electronic states at molecule/electrode interfaces are crucial factors for determining the transport and optoelectronic properties of molecular junctions.
Strong light--matter coupling offers a potential for manipulating these factors, enabling to boost in the efficiency and versatility of these junctions.
Here, we investigate electroluminescence from single-molecule junctions in which the molecule is strongly coupled with the vacuum electromagnetic field in a plasmonic nanocavity.
We demonstrate an improvement in the electroluminescence efficiency by employing the strong light--matter coupling in conjunction with the characteristic feature of single-molecule junctions to selectively control the formation of the lowest-energy excited state.
The mechanism of efficiency improvement is discussed based on the energetic position and composition of the formed polaritonic states.
Our findings indicate the possibility to manipulate optoelectronic conversion in molecular junctions by strong light--matter coupling and contribute to providing design principles for developing efficient molecular optoelectronic devices.
\end{abstract}


Optoelectronic conversion in organic materials has been a topic of great interest owing to their fundamental importance and potential applications in molecular devices~\cite{Burroughes1990, Sariciftci1992, Scholes2017, Imai-Imada2022}.
In light emission responding to the passage of an electric current, which is called electroluminescence, excitons are created through a combination of injected electrons and holes, and subsequently light emission occurs associated with the radiative decay of excisions.
Elucidation of molecular electroluminescence processes will provide knowledge for exciton and charge carrier dynamics in electronically pumped organic materials.
Moreover, it will aid in designing high-performance molecular optoelectronic devices such as organic light emitting diodes (OLEDs).

An efficient generation of emissive excited electronic states by charge injection is one of the most important factors for realizing the high-efficient OLEDs.
When uncorrelated charges are injected from electrodes into a molecular assembly and carriers with opposite charges merge on a molecule, excited electronic states with singlet and triplet spin multiplicity are formed in a 1:3 ratio~\cite{Baldo1998}.
The first excited electronic state with triplet spin multiplicity, ${\rm T}_1$ typically has lower energy than the singlet state, ${\rm S}_1$.
Compared with the emissive ${\rm S}_1$ state, the ${\rm T}_1$ state has significantly low radiative efficiencies because of the slow phosphorescent transition to the ${\rm S}_0$ state.
Therefore, the formation of triplet states can be a source of energy loss in these devices.
One route to tackle this issue is to harness the formed ${\rm T}_1$ state.
Efficient phosphorescent molecules~\cite{Baldo1998} and organic materials that harvest ${\rm T}_1$ states by converting them into ${\rm S}_1$ states~\cite{Uoyama2012, Singh-Rachford2010} have been developed for this purpose.
Another route to improve the electroluminescence efficiency could be to prevent the formation of ${\rm T}_1$ states and directly excite only luminescent electronic states.
To investigate the potential of this option, we focus on two areas: single-molecule junctions and the strong coupling (SC) regime of the light--matter interaction.

In a single-molecule junction, an electron and hole are injected directly from electrodes into a molecule, allowing the energy of the injected charges to be finely tuned by the bias voltage applied between the electrodes.
Taking advantage of this property, Kimura {\it et al.}~\cite{Kimura2019} succeeded experimentally in selectively forming the lowest-energy excited electronic state, the ${\rm T}_1$ state in a single molecule.
What should be noticed here is the difference from the formation mechanism of the ${\rm S}_1$ and ${\rm T}_1$ states in bulk systems.
In an electrically pumped molecular assembly, in general, both the states would be inevitably generated via uncontrollable diffusion and combination of the injected charges.

In the SC regime of the light--matter interaction, photoactive excitations in materials hybridize with the zero-point quantum fluctuations of the electromagnetic field, namely the vacuum field.
This leads to a formation of coupled light--matter states, which are termed polaritonic states.
The SC regime offers enormous potential for tailoring material properties~\cite{FriskKockum2019, Basov2020, Hubener2021, Bloch2022}, and hence, the regime has been investigated for use in various types of materials~\cite{Kaluzny1983, Weisbuch1992, Lidzey1998}.
Specifically, the SC of organic molecules to the vacuum field has been attracting attention as a novel tool with potential applications in many research areas including chemistry, physics, and materials sciences~\cite{Ebbesen2016, Garcia-Vidal2021, Ribeiro2018b, Hutchison2012, Herrera2016, Galego2016, Munkhbat2018b, Thomas2019a, Campos-Gonzalez-Angulo2019, Phuc2020, Orgiu2015a, Hutchison2013, Zhong2016a, Rozenman2018, Du2018d, Polak2020, Takahashi2019, Stranius2018, Yu2021, Eizner2019, Martinez-Martinez2019, Gubbin2014, Genco2018, Dunkelberger2016, Li2022b, Kena-Cohen2010, Keeling2020}.
Typically transitions between electronic states of different spin multiplicities are less photoactive, and hence, the singlet excited states mainly contribute to the formation of the polaritonic states.
On the basis of this characteristic, it was experimentally demonstrated that the energy levels of the polaritonic states composed of emissive ${\rm S}_1$ state and the vacuum field could be selectively modified without significantly disturbing the ${\rm T}_1$ state~\cite{Stranius2018, Yu2021, Eizner2019}.
Furthermore, the SC regime was realized for single molecules in a plasmonic nanocavity~\cite{Chikkaraddy2016}.
In this system, single molecules are strongly coupled with the zero-point quantum fluctuations of the electromagnetic field associated with the plasmon excitations, which is coined as the vacuum plasmonic field.

\begin{figure*}
\includegraphics[bb=0 0 510 150]{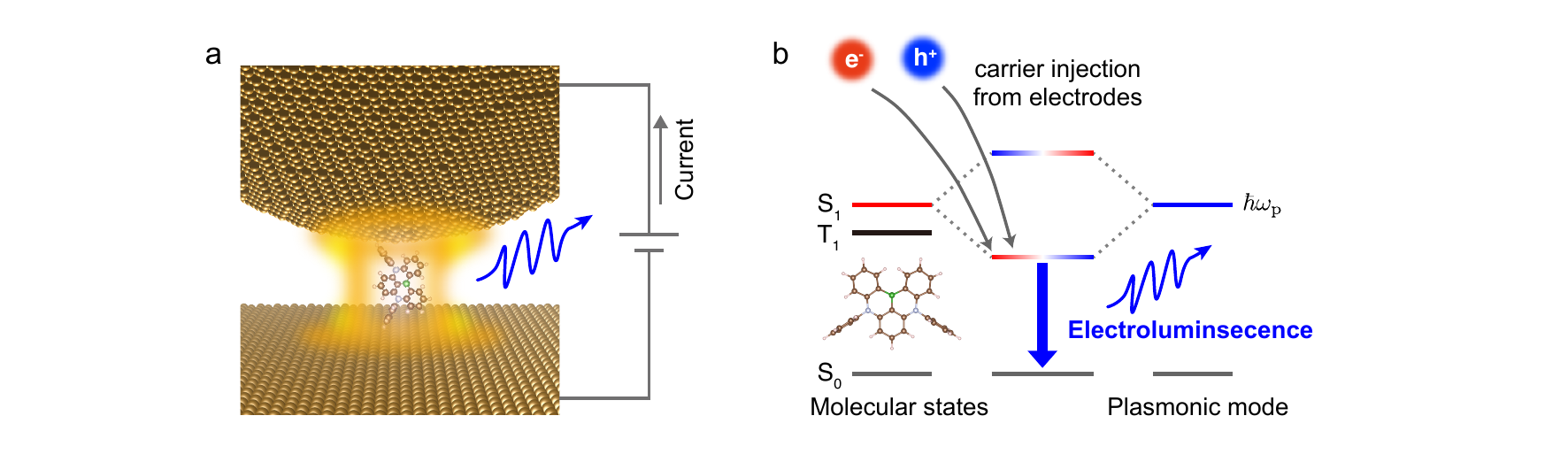}
\caption{
Single-molecule electroluminescence in the strong light--matter coupling regime.
(a) Illustration of a dye molecule in a plasmonic nanocavity. The molecule is excited by the carrier injection, and subsequently, photons are emitted from the polaritonic state composed of the fluorescent singlet state of the molecule and vacuum plasmonic field.
(b) Energy levels of molecular electronic states, plasmonic mode, and coupled polaritonic modes.
When the first excited polaritonic state is positioned lower than molecular triplet states in energy, the selective excitation of the polaritonic state without producing the harmful triplet states is achievable by carrier injection from electrodes, resulting in efficient electroluminescence.
}
\label{fig:Concept}
\end{figure*}

As aforementioned, the use of single-molecule junctions makes it possible to selectively control the formation of only the lowest-energy excited state, whereas the SC of molecules to the vacuum field can modify a desired luminescent state to be the lowest-energy excited state through the formation of polaritonic states.
Therefore, it is anticipated that the simultaneous use of both single-molecule junctions and the SC regime will circumvent the formation of the undesirable ${\rm T}_1$ state and selectively populate only luminescent states in a controllable manner.
In this study, we theoretically investigate electroluminescence from a single-molecule junction in a plasmonic nanocavity where a molecule is strongly coupled to the vacuum plasmonic field, as shown in Figure~\ref{fig:Concept}.
In the model, excited electronic states of a molecule are generated upon the electric injection of an electron and hole from the electrodes, and photons are emitted from the polaritonic state composed of the fluorescent singlet state and vacuum plasmonic field.
We resort to an open quantum system approach to explore electroluminescence under the influence of the surrounding environment composed of the radiation field and the intramolecular vibrations.
Electric current and photon flux are analyzed to clarify the effects of the polariton formation on carrier injection and energy conversion for a single molecule in the SC regime.
We demonstrate that the electroluminescence efficiency in a single-molecule junction can be effectively controlled by the degree of exciton--plasmon coupling strength and energy detuning.

We consider a junction consisting of a single molecule in a plasmonic nanocavity coupled to the radiation field and the bosonic bath that corresponds to molecular vibrations, electron--hole pairs and phonons in electrodes, and other excitations.
The molecule is in contact with two electrodes, leading to an electron transfer between the molecule and electrodes.
We construct a model Hamiltonian as $\hat{H} = \hat{H}_\mathrm{sys} + \hat{H}_\mathrm{env} + \hat{V}$.
In the equation, $\hat{H}_\mathrm{sys}$ is the Hamiltonian to describe the system composed of the molecular degrees of freedom (DOFs), $\hat{H}_\mathrm{M}$ and the local surface plasmons (LSPs) in the nanocavity, $\hat{H}_\mathrm{P}$, 
\begin{align}
	\hat{H}_\mathrm{sys} &= \hat{H}_\mathrm{M} + \hat{H}_\mathrm{P} + \hat{V}_\mathrm{M, P}
\label{eq:Hsys} 
\end{align}
with $\hat{V}_\mathrm{M, P}$ being the coupling between the molecular DOFs and the LSPs.
Diagonalizing Eq.~\eqref{eq:Hsys} yields polaritonic states.
We assume that the electrodes L and R, radiation field, and bosonic bath weakly interact with the system and regard such DOFs as the surrounding environment described by the Hamiltonian,
\begin{align}
	\hat{H}_\mathrm{env} 
    = 
    \sum_{K={\rm L,R}} \hat{H}_{K} + \hat{H}_\mathrm{rad} + \hat{H}_\mathrm{B},
    \label{eq:Henv} 
\end{align}
where $\hat{H}_{K}$, $\hat{H}_\mathrm{rad}$, and $\hat{H}_\mathrm{B}$ describe the electrode, radiation field, and bosonic bath, respectively.
The interactions between the subsystems are represented by
\begin{align}
	\hat{V} 
    = 
    \sum_{K={\rm L,R}} \hat{V}_{\mathrm{M}, K} + \hat{V}_\mathrm{P, rad} + \hat{V}_\mathrm{M, B} + \hat{V}_\mathrm{P, B}.
    \label{eq:Vsys-env}
\end{align}
In Eq.~\eqref{eq:Vsys-env}, $\hat{V}_{\mathrm{M},K}$ drives electron transfer between the molecule and the electrode $K$ characterized by the rate, $\Gamma_K$.
We assume that $\Gamma_K$ is independent of electronic states for simplicity~\cite{Fu2018, Miwa2019a}.
When a bias voltage $V$ is applied between the two electrodes, L and R, the chemical potential of the conduction electrons in the electrode $K(={\rm L, R})$ is expressed as $\mu_K = \mu_0 + \zeta_K e V$ with an adjustable parameter, $\zeta_K$.
For simplicity, we assume an asymmetric case of $\zeta_{\rm L} = 0$ and $\zeta_{\rm R} = -1$.
The second term in Eq.~\eqref{eq:Vsys-env}, $\hat{V}_\mathrm{P, rad}$ describes the coupling between an LSP and a photon in the radiation field \cite{Shishkov2016,Miwa2021}, inducing decay and dephasing of the LSP with time constants of $T_\mathrm{1,p}$ and $T_\mathrm{2,p}$, respectively.
The couplings $\hat{V}_\mathrm{M, B}$ and $ \hat{V}_\mathrm{P, B}$ induce the energy dissipation from the molecule and an LSP to the bosonic bath, yielding nonradiative decays of a molecular exciton and the LSP.
The rates of the non-radiative decays from the exciton and LSP are expressed as $\gamma_\mathrm{m,nr}$ and $\gamma_\mathrm{p,nr}$, respectively.
In this study, we trace out the environmental DOFs and thereby obtain an equation of motion for the reduced density operator that describes the strongly coupled molecular excitations and the LSPs under the weak influences of the electrodes, radiation field, and bosonic bath.
Specifically, the impacts of the rate constants $\Gamma_K$, $T_\mathrm{1,p}$, $T_\mathrm{2,p}$, $\gamma_\mathrm{m,nr}$, and $\gamma_\mathrm{p,nr}$ and the chemical potential $\mu_K$ can be investigated through the relaxation tensors that appears in the reduced density operator formalism.
Details on deriving the equation of motion for the reduced density operator are provided in the Supporting Information.

Explicit expressions of the individual Hamiltonians and the couplings are provided herein. 
Let $\ket{N,A}$ denote the electronic eigenstates of the isolated molecule, where $N$ represents the number of excess electrons with respect to the neutral molecule and $A$ labels the electronic state.
For the neutral molecule, $\ket{0,{\rm S}_1}$ denotes the first excited state with the singlet spin multiplicity, whereas $\ket{0, {\rm T}_1^{m=0, \pm1}}$ is the first excited state with the triplet spin multiplicity. The ground electronic states with the doublet spin multiplicity of the anion and cation are expressed as $\ket{+1,{\rm D}_0^{\sigma=\pm1/2}}$ and $\ket{-1,{\rm D}_0^{\sigma=\pm1/2}}$, respectively.
The energy of the $\ket{N,A}$ state is given by $E_{N,A}$. 
The optical response of plasmonic nanocavity can be modeled using a set of harmonic oscillators that describe the linear collective excitations of the electrons in a metal \cite{Shishkov2016, Miwa2021}.
To simplify the model, we consider a representative mode that dominantly contributes to the formation of polaritonic states and an optical response of the system.
Indeed, a single LSP mode was investigated to analyze the optical response of single molecules in the SC regime~\cite{Chikkaraddy2016}.
Therefore, we introduce the operators $\hat{a}_{\rm p}^\dagger$ and $\hat{a}_{\rm p}$ to create and annihilate an LSP with an energy $\hbar \omega_{\rm p}$ in the nanocavity, respectively.
In the system Hamiltonian in Eq.~\eqref{eq:Hsys}, $\hat{H}_\mathrm{M}$, $\hat{H}_\mathrm{P}$, and $\hat{V}_\mathrm{M, P}$ are expressed as 
\begin{align}
	\hat{H}_\mathrm{M} 
    &= \sum_{N,A} E_{N,A} \ket{N,A}\bra{N,A},
\\
	\hat{H}_\mathrm{P} 
    &= \hbar \omega_{\rm p} 
    \hat{a}^\dagger_{\rm p} \hat{a}_{\rm p} ,
\\
	\hat{V}_\mathrm{M, P}
    &= \hbar g ( \ket{0,{\rm S}_1}\bra{0,{\rm S}_0} + \hc )
    ( \hat{a}_{\rm p} + \hat{a}_{\rm p}^\dagger ),
\label{eq:VMC}
\end{align}
where $\hbar g$ is the coupling strength.
We consider only electronic transitions between the ground and first excited states that maintain the singlet spin multiplicity.
What should be emphasized here is that, as both the strong coupling and ultrastrong coupling regimes are investigated in this study, the rotating wave approximation is not employed~\cite{Cirio2016, Ridolfo2012, Beaudoin2011, DeLiberato2007, Braak2011}.
The SC of the $\ket{0,{\rm S}_0} - \ket{0,{\rm S}_1}$ transition to the plasmonic mode leads to the formation of the ground and $n$-th excited polaritonic states, $\ket{0, \mathrm{GS}}$ and $ \ket{0, {\rm P}_n}$, respectively.
For non-photoactive states such as $\ket{0,{\rm T}_1}$ and $\ket{\pm1, {\rm D}_0}$, an eigenstate of $\hat{H}_\mathrm{sys}$ is given as a tensor product form, $\ket{N,A; n_{\rm p}} = \ket{N,A}\otimes\ket{n_{\rm p}}$ with $\ket{n_{\rm p}}$ being an eigenstate of $\hat{H}_{\rm p}$ characterized by the number of plasmons, $n_{\rm p} = 0, 1, 2, \dots$.
In numerical calculations, we truncate the Hilbert space for plasmons to the low-lying four states.

As in the case of $\hat{H}_{\rm P}$, we introduce an operator $\hat{c}_{Kk\sigma}$ to annihilate an electron in a state $k$ of the electrode $K$ with an energy $\epsilon_{Kk\sigma}$ and a spin $\sigma$.
Further, we let the operators $\hat{b}_\alpha$ and $\hat{f}_\beta$ annihilate the $\alpha$-th photon in the radiation field with an energy $\hbar \omega_\alpha$ and the $\beta$-th bosonic mode in the bath with an energy $\hbar \omega_\beta$, respectively.
Their Hermitian conjugates indicate the respective creation operators.
Thus, the Hamiltonians in Eq.~\eqref{eq:Henv} are expressed as
$\hat{H}_{K} = \sum_{k \in K} \sum_{\sigma} \epsilon_{Kk\sigma} \hat{c}^\dagger_{Kk\sigma} \hat{c}_{Kk\sigma}$,
$\hat{H}_\mathrm{rad} = \sum_{\alpha} \hbar \omega_\alpha \hat{b}^\dagger_\alpha \hat{b}_\alpha$,
and $\hat{H}_\mathrm{B} = \sum_{\beta} \hbar \omega_\beta \hat{f}^\dagger_\beta \hat{f}_\beta$.
Furthermore, the Hamiltonians in Eq.~\eqref{eq:Vsys-env} are expressed as
\begin{align}
	\hat{V}_{\mathrm{M}, K} &= \sum_{N, A, B} \sum_{k \in K} \sum_{\sigma}
		( V_{Kk\sigma,N A B} \ket{N,A}\bra{N-1,B}\hat{c}_{Kk\sigma} + \hc  ) ,
\\
	\hat{V}_\mathrm{P, rad} &= \sum_{\alpha} U_\alpha
		( \hat{a}_{\rm p} + \hat{a}_{\rm p}^\dagger ) 
		( \hat{b}_\alpha + \hat{b}_\alpha^\dagger ),
        \label{eq:Vplasmon-rad}
\\
	\hat{V}_\mathrm{M, B} &= \sum_\beta \sum_{N, A, B} u^{\rm M}_{\beta, NAB}
		( \ket{N,A}\bra{N,B} + \mathrm{h.c} ) 
		( \hat{f}_\beta + \hat{f}_\beta^\dagger ),
\\
	\hat{V}_\mathrm{P, B} &= \sum_{\beta} u^{\rm p}_\beta
        ( \hat{a}_{\rm p} + \hat{a}_{\rm p}^\dagger ) 
		( \hat{f}_\beta + \hat{f}_\beta^\dagger ).
\end{align}
We assume that the spontaneous emission from a molecule is neglected because the coupling between molecular excitons and photons is sufficiently weaker than a plasmon--photon coupling.
Consequently, photons are expected to be emitted dominantly from LSPs.
This assumption is justified by experimental evidence that the time constant of the fluorescence decay of the molecule under investigation is on the order of nanoseconds~\cite{Hatakeyama2016}, whereas the decay and dephasing times of LSPs are typically on the order of several to several tens of femtoseconds~\cite{Maier2007, Sun2016}.
We note in passing that the molecule is coupled with an LSP through a longitudinal electric near-field generated with the LSP excitation~\cite{Shishkov2016, Miwa2021}.
It is known that this coupling enhances molecular optical transitions~\cite{Novotny2012, LeRu2009a}.

In case of weak molecule--electrode coupling, the steady state electric current across the molecule and electrode $K$ is expressed as~\cite{Fu2018}
\begin{equation}
    I_K 
    = 
    e \sum_{S,S_+,S_-} 
    ( k^K_{S_+ \gets S} - k^K_{S_- \gets S} ) 
    P_{S}
    \label{eq:current}
\end{equation}
where $e$ is the elementary charge, $P_S$ is the population of the system's eigenstate $\ket{S}$, and $k^{K}_{S' \gets S}$ denotes the transition rate from one state $\ket{S}$ to another $\ket{S'}$.
It is also noted that $\ket{S_\pm}$ represents the state in which the number of electrons is $\pm 1$ more than that of $\ket{S}$.
To consider the luminescence intensity, we calculate the photon flux $J_\mathrm{ph}$ from the molecule to the radiation field.
The photon flux is defined as the rate of change in the population of photons, $J_\mathrm{ph} = (d/dt) \sum_{\alpha} \langle \hat{b}_\alpha^\dagger(t) \hat{b}_\alpha(t) \rangle$.
The electroluminescence efficiency $\eta$ is obtained as $\eta = J_\mathrm{ph} e / \lvert I_K \rvert $.

We use 5,9-diphenyl-5,9-diaza-13b-boranaphtho[3,2,1-\textit{de}]anthracene (DABNA-1) \cite{Hatakeyama2016, MadayanadSuresh2020, Shizu2022, Pratik2022}, which exhibits a small singlet-triplet energy gap, $\Delta E_{\rm ST} < 0.20~\mathrm{eV}$ \cite{Hatakeyama2016} and relatively large oscillator strength, $f_\mathrm{osc} \simeq 0.31$ \cite{Pershin2019, Hall2020}, as the molecule for the junction.
The eigenenergies of the isolated molecule $\{E_{N,A}\}$ are determined with the previously reported values of the excitation energies, ionization potential, and electron affinity~\cite{Hatakeyama2016}.
The implications of the image interaction with the electrodes on the eigenenergies~\cite{Neaton2006, Perrin2013b, Fu2018, Imai-Imada2018, Miwa2019} are included in the values of $\{E_{N,A}\}$.

\begin{table}
  \caption{Parameters utilized in numerical calculations}
  \label{tbl:Parameters}
  \begin{tabular}{L{15truemm}R{25truemm} C{15truemm} L{15truemm}R{25truemm}}
    \hline
    parameter	& value	&& parameter	& value\\
    \hline
    $E_{0,{\rm S}_0}$		& $0~{\rm eV}$		    && $T_{\rm 1, p}$			& $5~{\rm fs}$          \\
    $E_{0,{\rm S}_1}$		& $2.67~{\rm eV}$	    && $T_{\rm 2, p}$			& $10~{\rm fs}$          \\
    $E_{0,{\rm T}_1}$		& $2.49~{\rm eV}$       && $\hbar\gamma_\mathrm{p,nr}$	& $10^{-4}~{\rm eV}$  \\
    $E_{-1,{\rm D}_0}$      & $5.58~{\rm eV}$       && $\hbar\Gamma_\mathrm{L}$		& $10^{-4}~{\rm eV}$  \\
    $E_{+1,{\rm D}_0}$      & $-2.91~{\rm eV}$	&& $\hbar\Gamma_\mathrm{R}$	    & $10^{-4}~{\rm eV}$  \\
    $\mu_0$                 & $-2.5~{\rm eV}$		&& $\hbar\gamma_\mathrm{m,nr}$	& $8.0 \times 10^{-9}~{\rm eV}$ \\
    $T$			& $300~{\rm K}$		& \\
    \hline
  \end{tabular}
\end{table}

The parameters employed are listed in Table~\ref{tbl:Parameters}.
We adapt $\mu_0 = -2.5~\mathrm{eV}$ as the chemical potential of the conduction electrons in the electrodes at the zero bias voltage.
This value corresponds to the work function of the LiF film-covered Al electrode~\cite{Shaheen1998}, which was utilized in the device manufactured in Ref.~\citenum{Hatakeyama2016}.
Within the bias voltage range under investigation, we only consider $\ket{0,{\rm S_0}}$, $\ket{0,{\rm S_1}}$, $\ket{0, {\rm T}_1^{m=0, \pm 1}}$, $\ket{+1,{\rm D}_0^{\sigma=\pm 1/2}}$, and $\ket{-1,{\rm D}_0^{\sigma=\pm 1/2}}$.
Other necessary parameters are determined to be representative of realistic experimental situations. 
Unless otherwise stated, the temperature is set to 300~K. 
The exciton--plasmon coupling strength $\hbar g$ is chosen from several tens to several hundred milli-electron volts because similar values were reported in the experimental and computational studies~\cite{Chikkaraddy2016, Rossi2019, Kuisma2022}.
The decay and dephasing times of LSPs are set to $T_\mathrm{1,p} = 5~{\rm fs}$ and $T_\mathrm{2,p} = 10~{\rm fs}$, respectively~\cite{Maier2007}.
The non-radiative decay rate of LSPs is set as $\hbar\gamma_\mathrm{p,nr} = 10^{-4}~\mathrm{eV}$, so that the radiative decay of LSPs is dominant over the nonradiative one.
It is known that the radiative decay pathway can dominate for LSPs in certain kinds of metallic nanostructures such as aluminium nanospheres~\cite{Miwa2021}.
Because molecule--electrode couplings are weakened in single-molecule electroluminescence experiments~\cite{Qiu2003, Zhang2016, Imada2016, Doppagne2017, Kuhnke2017}, we assume a small value of coupling strength $\hbar\Gamma_K = 10^{-4}~\mathrm{eV}$.
The nonradiative decay rate of molecular exciton is set as $\hbar\gamma_{\rm m, nr} = 8 \times 10^{-9}~\mathrm{eV}$ to render its order similar to that reported previously~\cite{Shizu2022}.
For simplicity, spontaneous emission from the molecule, intersystem crossing (ISC), and reverse ISC processes are neglected in this study.
This is justified by the reports that these processes are significantly slower than the radiative decay of LSPs and the electron transfer between the molecule and electrodes~\cite{Shizu2022, Stranius2018, Eizner2019, Martinez-Martinez2019, Yu2021}.
Therefore, it is reasonable to state that the dominant processes in electroluminescence is considered in the proposed scheme.


\begin{figure}
\includegraphics[bb=0 0 249 368]{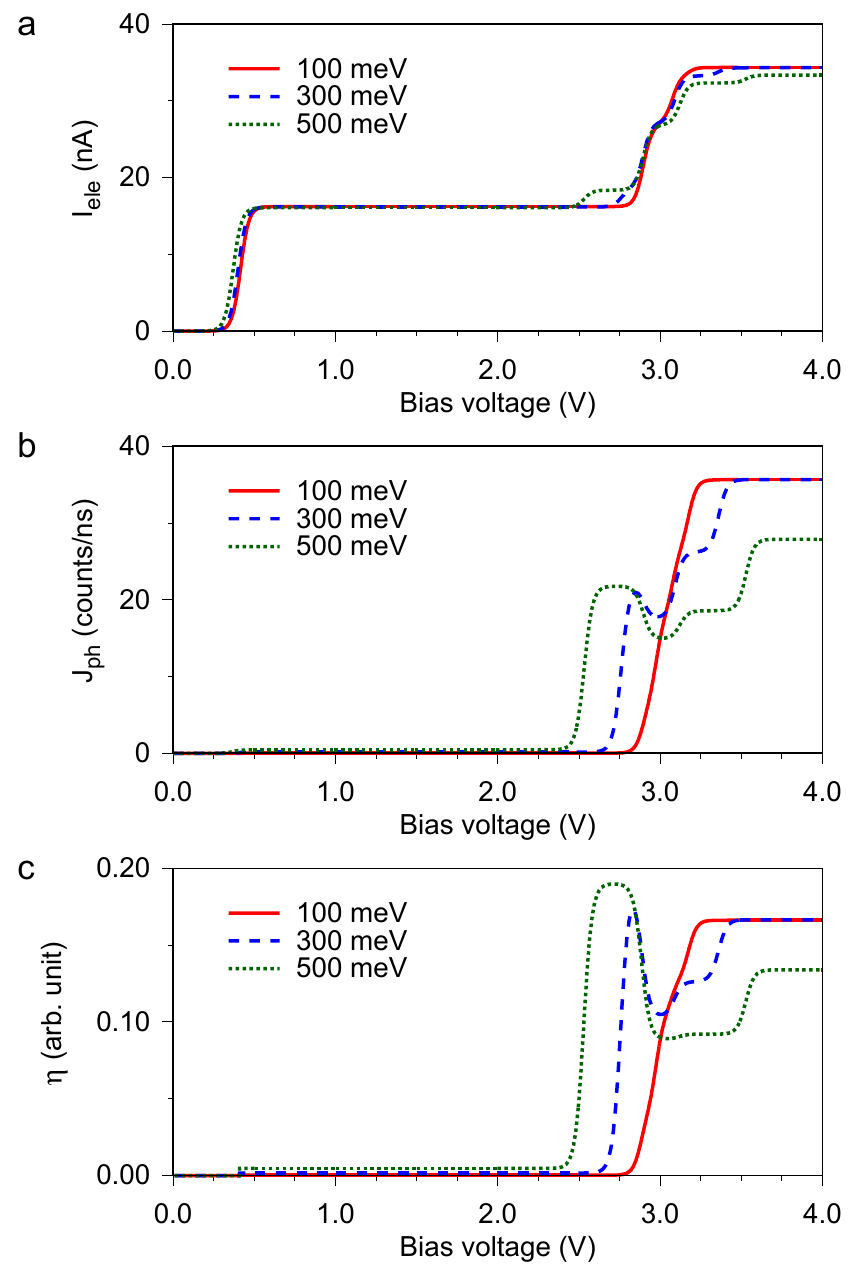}
\caption{
Bias voltage dependence of the observables.
Dependence of electric current $I_\mathrm{L}$, photon flux $J_\mathrm{ph}$, and electroluminescence efficiency $\eta$ on the bias voltage $V$ at the plasmon energy resonant with the excitation energy from the ground to first excited electronic states with the singlet spin multiplicity of the molecule $\hbar\omega_{\rm p} = E_{0, {\rm S_1}}$.
Red solid, blue dashed, and green dotted lines represent the results for the exciton--plasmon coupling strength $\hbar g = 100$, $300$, and $500~\mathrm{meV}$, respectively.}
\label{fig:Reso}
\end{figure}

Figure~\ref{fig:Reso} shows the numerical results of the electric current $I_\mathrm{ele}$, photon flux $J_\mathrm{ph}$, and electroluminescence efficiency $\eta$ as a function of the bias voltage $V$ for various values of the exciton--plasmon coupling strength.
The plasmon energy is chosen to be resonant with the $\ket{0,{\rm S_0}} - \ket{0,{\rm S_1}}$ transition so that $\hbar \omega_{\rm p} = E_{0, {\rm S_1}}$.

First, we focus on the current-voltage characteristic curves shown in Fig.~\ref{fig:Reso}a.
In the case of $V = 0~{\rm V}$, the molecule and plasmon are in the state $\ket{+1,{\rm D}_0}$ and $\ket{0}$, respectively.
The rises in the vicinity of $V = 0.25~{\rm V}$ and $2.75~\mathrm{V}$ indicate the onsets of the electron transfer associated with the transitions from $\ket{+1, {\rm D}_0; 0}$ to $\ket{0, \mathrm{GS} }$ and $\ket{0, {\rm T}_1; 0}$, respectively.
The transition from $\ket{+1, {\rm D}_0; 0}$ to the first excited polaritonic state $\ket{0, {\rm P}_1}$ is turned on near $V = 2.80$, $2.60$, and $2.40~\mathrm{V}$ in the cases of $\hbar g = 100$, $300$, and $500~\mathrm{meV}$, respectively.
Figure~S2 of SI plots the $\ket{0, {\rm P_1}}$ state energy, $E_{\rm P_1}$, as a function of $\hbar g$.
In what follows, the threshold voltages required for the transitions to $\ket{0, {\rm T_1}; 0}$ and $\ket{0, {\rm P_1}}$ are denoted by $V_{\rm T_1}$ and $V_{\rm P_1}$, respectively.

Next, we analyze the bias voltage dependence of the photon flux $J_\mathrm{ph}$ and the luminescence efficiency $\eta$.
Figure~\ref{fig:Reso}b shows that $J_\mathrm{ph}$ significantly increases near $V = 2.80$, $2.60$, and $2.40~\mathrm{V}$ for $\hbar g = 100$, $300$, and $500~\mathrm{meV}$, respectively.
As shown in Figure~S2 of SI, the energetic position of $E_{\rm P_1}$ is lower than that of the triplet excited state energy $E_{0,{\rm T_1}}$ when the coupling strength $\hbar g$ exceeds 175~meV.
Hence, for $\hbar g \ge 175~{\rm meV}$, the selective excitation of $\ket{0, {\rm P_1}}$ without populating $\ket{0,{\rm T}_1; 0}$ is realized in the range of the bias voltage, $V_{\rm P_1} < V < V_{\rm T_1}$.
For relatively large coupling strength such as $\hbar g \ge 300~{\rm meV}$, $J_\mathrm{ph}$ decreases near $V = 2.75~{\rm V}$.
At this bias voltage, the state $\ket{0, {\rm T}_1; 0}$ is populated, and therefore, the population of $\ket{0, {\rm P}_1}$ and consequently $J_\mathrm{ph}$ decrease.
The maximum value of $J_\mathrm{ph}$ is reached in the regions of $V \ge 3.30$, $3.50$, and $3.70~\mathrm{V}$ in the case of $\hbar g = 100$, $300$, and $500~\mathrm{meV}$, respectively.
For the case of $\hbar g = 100~{\rm meV}$, $\eta$ has the maximum value of 0.166 in this range of $V$ (Figure~\ref{fig:Reso}c).
For $\hbar g = 300~{\rm meV}$, the maximum value of electroluminescence efficiency $\eta_\mathrm{max}$ is 0.173 obtained at $V \sim 2.83~{\rm V}$.
For $\hbar g = 500~{\rm meV}$, where higher values of $\eta$ are obtained in the range of $V_{{\rm P}_1} < V < V_{{\rm T}_1}$, $\eta_\mathrm{max}$ is 0.190.
The results suggest that the strong light--matter coupling enables the selective excitation of $\ket{0, {\rm P}_1}$ by carrier injection without populating the triplet states, improving electroluminescence efficiency.

\begin{figure}
\includegraphics[bb=0 0 249 368]{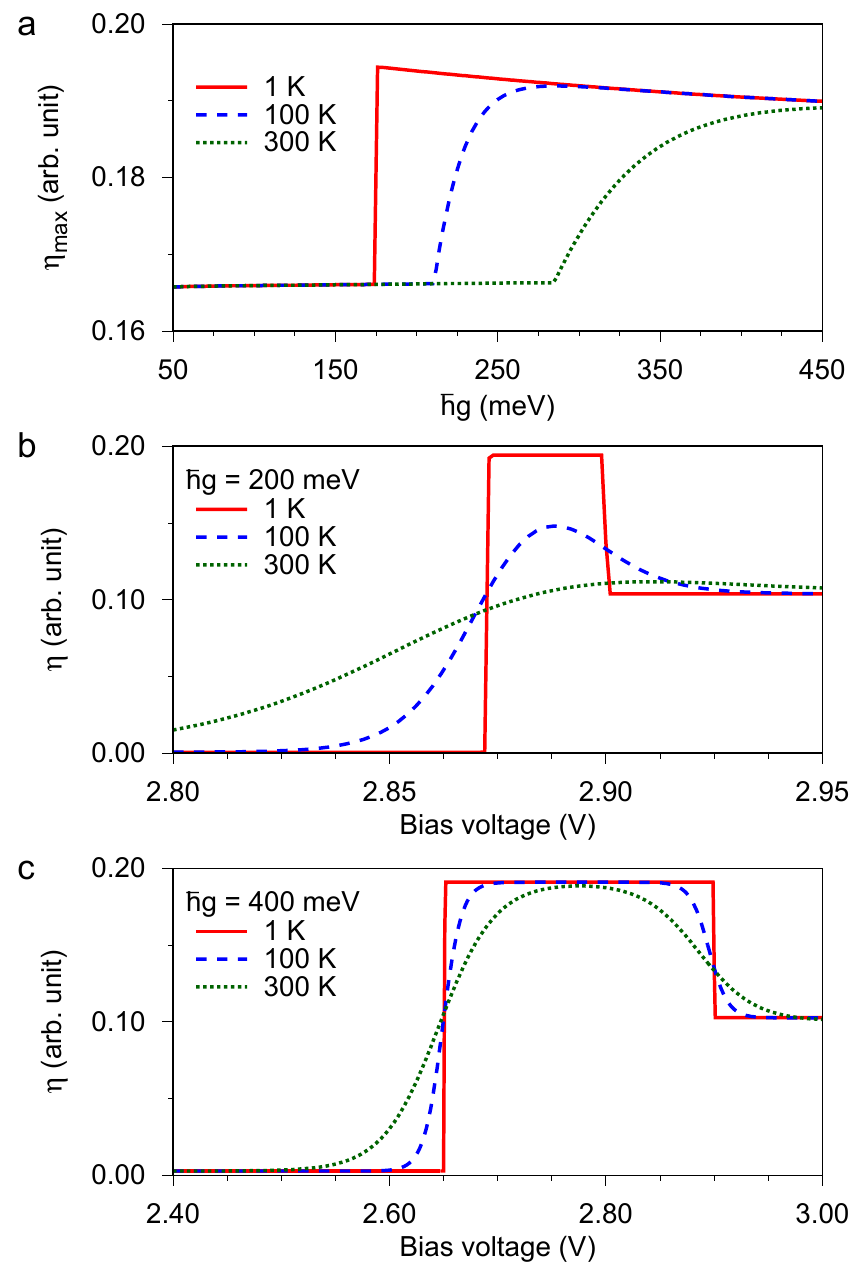}
\caption{
Electroluminescence efficiency in the resonance case.
(a) Plot of the maximum values of electroluminescence efficiency $\eta$ as a function of the exciton--plasmon coupling strength, $\hbar g$.
The bias voltage $V$ is applied in the range from 0~V to 5~V.
The dependence of $\eta$ on $V$ in the cases of (b) $\hbar g = 200$ and (c) $\hbar g = 400~{\rm meV}$.
Red solid, blue dashed, and green dotted lines represent the results for temperatures, $T = 1$, $100$, and $300~\mathrm{K}$, respectively.}
\label{fig:Efficiency}
\end{figure}

Figure~\ref{fig:Efficiency}a plots the maximum value of electroluminescence efficiency as a function of $\hbar g$ for various temperatures when the bias voltage in the range of $0<V<5\,{\rm V}$ is applied.
At low temperature $T = 1~{\rm K}$, $\eta_\mathrm{max}$ takes the value of 0.166 independently of $\hbar g$ when $\hbar g < 175~{\rm meV}$ is considered.
However, $\eta_\mathrm{max}$ takes a higher value, 0.194 when the coupling strength $\hbar g$ exceeds 175~meV, where $E_{{\rm P}_1}$ is lower than $E_{0,{\rm T}_1}$.
Figure~\ref{fig:Efficiency}b shows the bias voltage dependence of $\eta$ when $\hbar g = 200~{\rm meV}$ is considered.
As temperature increases, the Fermi distribution function in the vicinity of the Fermi level becomes smeared out, and $\ket{0, {\rm P}_1}$ is not sufficiently populated in the range of $V_{{\rm P}_1} < V < V_{{\rm T}_1}$.
The smearing of the Fermi distribution function arises over an energy range of the order of the thermal energy.
Hence, for the higher temperature case, a higher value of $\hbar g$ is needed to obtain $\eta_\mathrm{max}$ in the range of $V_{{\rm P}_1} < V < V_{{\rm T}_1}$.
Figure~\ref{fig:Efficiency}c shows the bias voltage dependence of $\eta$ when $\hbar g = 400~{\rm meV}$ is considered.
Because the interval between $V_{{\rm P}_1}$ and $V_{{\rm T}_1}$ is widened, the maximum value of $\eta$ can be reached in the range of $V_{{\rm P}_1} < V < V_{{\rm T}_1}$ even at high temperature.

\begin{figure*}
\includegraphics[bb=0 0 510 288]{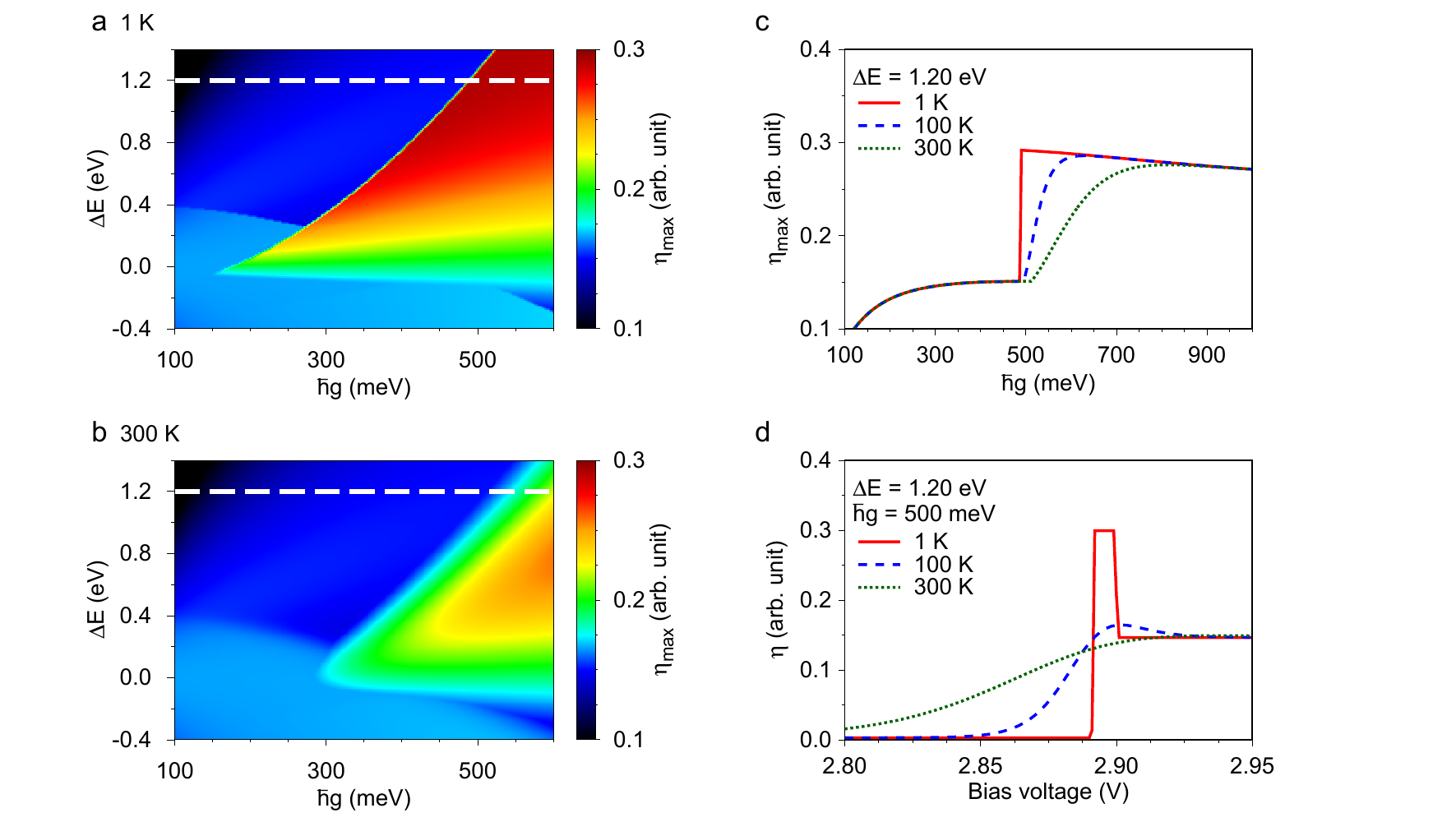}
\caption{
Dependence of electroluminescence efficiency on the energy detuning and coupling strength.
Two dimensional plot of the maximum values of electroluminescnce efficiency $\eta$ as a function of the exciton--plasmon coupling strength $\hbar g$ and energy detuning $\Delta E$ for temperatures of (a) $T = 1~\mathrm{K}$ and (b) $T = 300~\mathrm{K}$.
The bias voltage $V$ is applied in the range from 0~V to 5~V.
(c) Plot of the maximum values of $\eta$ as a function of $\hbar g$ for $\Delta E = 1.20~\mathrm{eV}$.
(d) The dependence of $\eta$ on $V$ for $\Delta E = 1.20~{\rm eV}$ and $\hbar g = 500~{\rm meV}$.
Red solid, blue dashed, and green dotted lines represent the results for $T = 1$, $100$, and $300~\mathrm{K}$, respectively.
}
\label{fig:2Defficiency}
\end{figure*}

For further improvement in electroluminescence efficiency, the effects of the energy detuning $\Delta E = \hbar\omega_{\rm p} - E_{0, {\rm S_1}}$ are analyzed.
Figures~\ref{fig:2Defficiency}a and \ref{fig:2Defficiency}b show two-dimensional plots of $\eta_\mathrm{max}$ as a function of $\Delta E$ and $\hbar g$ for temperatures $T = 1~{\rm K}$ and $300~\mathrm{K}$, respectively.
At low temperature $T = 1~{\rm K}$, $\eta_\mathrm{max}$ takes the highest value, 0.292 at $\Delta E \sim 1.20~{\rm eV}$ and $\hbar g \sim 490~{\rm meV}$.
Dependence of $E_{{\rm P}_1}$ on $\hbar g$ in the case of $\Delta E = 1.20~{\rm eV}$ is shown in Figure~S3 of SI.
The energetic position of $E_{{\rm P}_1}$ becomes lower than that of $E_{0,{\rm T}_1}$ when $\hbar g$ exceeds 488~meV.
Figure~\ref{fig:2Defficiency}c plots $\eta_\mathrm{max}$ as a function of $\hbar g$ in the case of $\Delta E = 1.20~{\rm eV}$.
The maximum value of electroluminescence efficiency $\eta_\mathrm{max}$ drastically increases when $\hbar g$ exceeds 488~meV.
Figure~\ref{fig:2Defficiency}d shows the bias voltage dependence of $\eta$ in the case of $\Delta E = 1.20~{\rm eV}$ and $\hbar g = 500~{\rm meV}$.
In this parameter setting, the threshold voltages required for the transitions to $\ket{0, {\rm T}_1; 0}$ and $\ket{0, {\rm P}_1}$ are approximately 2.90~V and 2.89~V, respectively.
At $T = 1~{\rm K}$, the luminescence efficiency $\eta$ takes its maximum value in the range of $V_{{\rm P}_1} < V < V_{{\rm T}_1}$.
The results indicate that the selective excitation of $\ket{0, {\rm P}_1}$ realizes an efficient electroluminescence also in the nonresonant case.

Figures~\ref{fig:2Defficiency}b shows that $\eta_\mathrm{max}$ for $T = 300~{\rm K}$ has its highest value in the case of a relatively larger coupling strength $\hbar g$ in comparison to that for $T = 1~{\rm K}$.
This is owing to the smearing of the Fermi distribution function.
For a verification purpose, we analyze the bias voltage dependence of $\eta$ in the case of $T = 300~\mathrm{K}$, $\Delta E = 1.20~\mathrm{eV}$, and $\hbar g = 500~\mathrm{meV}$ shown in Figure~\ref{fig:2Defficiency}d.
In this case, the interval between $V_{{\rm P}_1}$ and $V_{{\rm T}_1}$ is as narrow as approximately 0.01~V.
The corresponding energy, 0.01~eV, is small in comparison to the thermal energy at $ T = 300~{\rm K}$.
Therefore, $\ket{0, {\rm P}_1}$ is not sufficiently populated in the range of $V_{{\rm P}_1} < V < V_{{\rm T}_1}$ at this temperature.

The numerical results has demonstrated that the selective excitation of $\ket{0, {\rm P}_1}$ without populating $\ket{0,{\rm T}_1; 0}$ provides efficient electroluminescence.
To gain an insight into the obtained results and to explore possible optimal setups for realizing efficient electroluminescence, we derive an analytic expression for the efficiency $\eta$ by solving the rate equations to describe time-evolution of populations of eigenstates of the system Hamiltonian, Eqs.~(S29)-(S31) of SI.
For simplicity, we employ the following assumptions: symmetric molecule--electrode couplings $\Gamma_\mathrm{L} = \Gamma_\mathrm{R} = \Gamma$ and zero temperature $T = 0~{\rm K}$.
We further assume that the following two conditions are satisfied: the coupling strength $\hbar g$ is strong enough to satisfy $E_{{\rm P}_1} \le E_{0,{\rm T}_1}$, and the bias voltage is in the region of $V_{\rm P_1} < V < V_{\rm T_1}$.
Under these assumptions, the analytic expression is obtained as
\begin{align}
    \eta
    & = 
    \frac{ \lvert c_{{\rm S}_1}^{{\rm P}_1} \rvert^2 }
         { \lvert c_{{\rm S}_1}^{{\rm P}_1} \rvert^2+ 2 \lvert c_{{\rm S}_0}^{\mathrm{GS}} \rvert^2 }
    \frac{ \kappa ( k_\mathrm{p, r} + k_\mathrm{p, nr} ) }
         { \kappa ( k_\mathrm{p, r} + k_\mathrm{p, nr} ) + 2 \Gamma \lvert c_{{\rm S}_1}^{{\rm P}_1} \rvert^2 }
    \label{eq:eta1}
\end{align}
with $c_{{\rm S}_1}^{{\rm P}_1} = \left\langle 0, {\rm S}_1; 0 \middle\vert 0, {\rm P}_1 \right\rangle$,
$c_{{\rm S}_0}^{{\rm GS}} = \left\langle 0, {\rm S}_0; 1 \middle\vert 0, {\rm GS} \right\rangle$, and
$\kappa = \lvert \left\langle 0, {\rm P}_1 \middle\vert \left( \hat{a}_{\rm p} + \hat{a}_{\rm p}^\dagger \right) \middle\vert 0, \mathrm{GS} \right\rangle \rvert^2$.
Rates of the radiative and nonradiative decays of the first excited polaritonic state are given by $\kappa k_\mathrm{p, r}$ and $\kappa k_\mathrm{p, nr}$, respectively.
One can obtain an optimal setup that maximizes $\eta$ under the condition of $E_{{\rm P}_1} \le E_{0,{\rm T}_1}$.
Smaller values of $k_\mathrm{p, nr} / k_\mathrm{p, r}$ and $\Gamma / \kappa k_\mathrm{p, r}$ lead to a higher value of $\eta$.
Furthermore, the result indicates that the composition of the polaritonic state is important for determining the efficiency.
Larger values of $\lvert c_{{\rm S}_1}^{{\rm P}_1} \rvert^2$, i.e., more component of the molecular excited state in the first excited polaritonic state, leads to larger values of the first factor in the right-hand side of Eq.~(\ref{eq:eta1}), resulting in high efficiency for the blue detuning cases.
The radiative decay rate $\kappa k_\mathrm{p, r}$ decreases with an increase in $\Delta E$, and therefore, when the energy detuning exceeds a certain value, $\eta$ decreases with increasing $\Delta E$.

To simplify the analytic expression, the rotating wave approximation (RWA) is employed in the exciton--plasmon coupling in Eq.~\eqref{eq:VMC}.
The electroluminescence efficiency is recast as
\begin{align}
	\eta^\mathrm{RWA}
	&=
    \frac{2}{6+ (\sqrt{\alpha^2 + 4} - \alpha )^2 }
    \frac{ ( k_{\rm p, r} + k_{\rm p, nr} ) ( \sqrt{ \alpha^2 +4 } - \alpha )^2 }
    { ( k_{\rm p, r} + k_{\rm p, nr} ) ( \sqrt{ \alpha^2 +4 } - \alpha )^2 + 8 \Gamma },
\label{eq:eta2}
\end{align}
where $\alpha = \Delta E / \hbar g$ has been introduced. 
In the limit of $k_\mathrm{p, nr} \to 0$ and $\Gamma \to 0$, the upper bound of the electroluminescence efficiency is obtained as $\eta^\mathrm{RWA} = 1/3$.
In the limit of $\alpha \to \infty$ with a finite value of $\Gamma$, the value of $\eta^\mathrm{RWA}$ vanishes.

For further analysis with concrete parameter sets, we assume $k_\mathrm{p, nr} = 0$ and
$k_\mathrm{p, r} = 2 E_{{\rm P}_1}^3/[T_\mathrm{2,p}(\hbar\omega_{\rm p})^3]$.
Moreover, the value of $\hbar g$ is set to satisfy $E_{\rm P_1} = E_{0, {\rm T_1}}$, because the efficiency is expected to be high near this parameter setting as demonstrated in the computational results.
Consequently, the efficiency is obtained as
\begin{equation}
	\eta^\mathrm{RWA} 
    =
	\frac{ \Delta E + \Delta E_{\rm ST} }{ 3 \Delta E + 5 \Delta E_{\rm ST} }
	\left[ 
        1 + 
        \gamma ( \Delta E + E_{0, {\rm S_1}} )^3
        ( \Delta E + \Delta E_{\rm ST} ) 
    \right]^{-1},
    \label{eq:eta3}
\end{equation}
where $ \Delta E_{\rm ST} = E_{0, {\rm S}_1} - E_{0, {\rm T}_1}$ and $\gamma = \Gamma T_{\rm 2,p} / ( E_{0, {\rm T}_1}^3 \Delta E_{\rm ST} )$ have been defined.
With the parameters of $E_{0, {\rm S}_1} = 2.67~\mathrm{eV}$, $E_{0, {\rm T}_1} = 2.49~\mathrm{eV}$, $\hbar \Gamma = 10^{-4}~\mathrm{eV}$, and $T_\mathrm{2,p} = 10~\mathrm{fs}$, we obtain $\eta^\mathrm{RWA} = 0.285$ as the maximum value of the luminescence efficiency near $\Delta E = 0.91~\mathrm{eV}$.
The results correspond to the values obtained with the numerical calculations.
It is noteworthy that the RWA in the exciton--plasmon coupling was not employed for the numerical calculations, and therefore, the system's eigenstates and eigenenergies are different from those utilized in the derivation of $\eta^\mathrm{RWA}$.
This leads to a quantitative difference between the analytical solution and the numerical result; nevertheless, qualitative trends are well captured.

In conclusion, we have demonstrated the efficient electroluminescence from a strongly coupled system of a single molecule in a plasmonic nanocavity.
The electroluminescence efficiency is significantly improved when the coupling strength is strong enough so that the energetic position of the first excited polaritonic state is lower than that of the triplet excited state of a molecule, where a selective excitation of the first excited polaritonic state without populating the triplet excited state can be implemented by an electric injection of carriers into a molecule in a certain bias voltage range.
The parameter set which maximizes the electroluminescence efficiency is determined through numerical calculations, sweeping the exciton--plasmon coupling strength and plasmonic energy.
Furthermore, an analytic expression of the electroluminescence efficiency, which is beneficial to understanding the general characteristics of the present model and to exploring an optimal setup for efficient electroluminescence concisely, is derived.
The results obtained in this study will contribute to understanding charge and energy flow in strongly coupled light--matter systems.
Moreover, they illustrate the potential to control the transport and optoelectronic properties of molecular junctions by engineering the zero-point quantum fluctuations of the surrounding electromagnetic field.


\begin{suppinfo}
The following file is available free of charge.
\begin{itemize}
  \item Supporting Information: Eigenstates of the isolated molecule, Diagonalization of the system Hamiltonian, Rate constants in the system's eigenstate representation, Derivation of quantum master equation, Formulation of photon flux, Derivation of analytic expression of the electroluminescence efficiency (PDF)
\end{itemize}
\end{suppinfo}

\begin{acknowledgement}
K.M. thanks Prof. Kiyoshi Miyata for fruitful discussions.
The molecular structure is visualized with the use of VESTA~\cite{Momma2011}.
This work was supported by JSPS KAKENHI (Grant Numbers JP21K14481 and JP21H01052) and MEXT Quantum Leap Flagship Program (Grant Numbers JPMXS0118069242 and JPMXS0120330644).
\end{acknowledgement}

\input{main.bbl}



\section*{Supplementary material}

# Eigenstates of the isolated molecule

In this study, we taken into account the following eigenstates of the isolated molecule written in the second quantized form:

$$|0, S_0\rangle = \hat{d}^\dagger_{H\uparrow}\hat{d}^\dagger_{H\downarrow}|\chi\rangle, \tag{S1}$$

$$|0, S_1\rangle = \frac{1}{\sqrt{2}}\left(\hat{d}^\dagger_{H\uparrow}\hat{d}^\dagger_{L\downarrow} - \hat{d}^\dagger_{H\downarrow}\hat{d}^\dagger_{L\uparrow}\right)|\chi\rangle, \tag{S2}$$

$$|0, T_1^0\rangle = \frac{1}{\sqrt{2}}\left(\hat{d}^\dagger_{H\uparrow}\hat{d}^\dagger_{L\downarrow} + \hat{d}^\dagger_{H\downarrow}\hat{d}^\dagger_{L\uparrow}\right)|\chi\rangle, \tag{S3}$$

$$|0, T_1^{+1}\rangle = \hat{d}^\dagger_{H\uparrow}\hat{d}^\dagger_{L\uparrow}|\chi\rangle, \tag{S4}$$

$$|0, T_1^{-1}\rangle = \hat{d}^\dagger_{H\downarrow}\hat{d}^\dagger_{L\downarrow}|\chi\rangle, \tag{S5}$$

$$|-1, D_0^{+\frac{1}{2}}\rangle = \hat{d}_{H\uparrow}|\chi\rangle, \tag{S6}$$

$$|-1, D_0^{-\frac{1}{2}}\rangle = \hat{d}_{H\downarrow}|\chi\rangle, \tag{S7}$$

$$|+1, D_0^{+\frac{1}{2}}\rangle = \hat{d}^\dagger_{H\uparrow}\hat{d}^\dagger_{H\downarrow}\hat{d}^\dagger_{L\uparrow}|\chi\rangle, \tag{S8}$$

$$|+1, D_0^{-\frac{1}{2}}\rangle = \hat{d}^\dagger_{H\uparrow}\hat{d}^\dagger_{H\downarrow}\hat{d}^\dagger_{L\downarrow}|\chi\rangle, \tag{S9}$$

where $d^\dagger_{m\sigma}$ is a creation operator for an electron on the molecular orbital $m$ with spin $\sigma$. The highest occupied molecular orbital (HOMO) and lowest unoccupied molecular orbital (LUMO) for the ground electronic state $S_0$ with spin singlet multiplicity of the neutral molecule are labeled as H and L, respectively. $|\chi\rangle$ represents a reference state in which both HOMO and LUMO are unoccupied.[1] We employ $|N, A\rangle$ to label the molecular eigenstates, where the definitions of $N$ and $A$ are given in the main text. We assume that molecular subspace is spanned by the above-shown nine eigenstates.

The rate constants for the transition between $|N, A\rangle$ and $|N-1, B\rangle$ with accompanying an electron transfer between the molecule and the electrode $K$ ($K = L, R$) are given by

$$\Gamma^K_{N,A,B} = \frac{2\pi}{\hbar}\sum_{k \in K}\sum_\sigma |V_{Kk\sigma, NAB}|^2 \delta(\hbar\omega - \epsilon_{Kk\sigma}), \tag{S10}$$

where $V_{Kk\sigma,NAB}$ and $\epsilon_{Kk\sigma}$ are, respectively, the molecule-lead coupling strength in the picture of molecular eigenstates and the energy of a conduction electron in state $k$ of the electrode $K$ with the spin $\sigma$. Hereafter, we make the following assumptions: the molecule-lead coupling strength is weak, the overlap integral of the molecular orbital $m$ and the conduction electron in the electrode $K$ is independent on $m$, and $\Gamma^K_{N,A,B}$ is energy-independent. Under these assumptions, the rate constants can be expressed as

$$\Gamma^K_{N,A,B} = \Gamma_K \nu_{N,A;N-1,B}, \tag{S11}$$

where $\Gamma_K$ is the constant characterizing the molecule-electrode coupling strength. Here, according to Ref. 1, we has introduced the dimensionless coupling coefficient $\nu_{N,A;N-1,B}$ of the transition between $|N,A\rangle$ and $|N-1,B\rangle$, which is given by

$$\nu_{N,A;N-1,B} = \left| \left\langle N, A \left| \sum_{m,\sigma} d^\dagger_{m,\sigma} \right| N-1, B \right\rangle \right|^2. \tag{S12}$$

The values of $\nu_{N,A;N-1,B}$ for the molecular eigenstates considered in this study are summarized in Table S1.

Table S1: Values of $\nu_{N,A;N-1,B}$ for the molecular eigenstates considered in this study

|  | $|-1, D_0^{+1/2}\rangle$ | $|-1, D_0^{-1/2}\rangle$ |  | $|+1, D_0^{+1/2}\rangle$ | $|+1, D_0^{-1/2}\rangle$ |
|---|---|---|---|---|---|
| $|0, S_0\rangle$ | 1 | 1 | $|0, S_0\rangle$ | 1 | 1 |
| $|0, S_1\rangle$ | $\frac{1}{2}$ | $\frac{1}{2}$ | $|0, S_1\rangle$ | $\frac{1}{2}$ | $\frac{1}{2}$ |
| $|0, T_1^0\rangle$ | $\frac{1}{2}$ | $\frac{1}{2}$ | $|0, T_1^0\rangle$ | $\frac{1}{2}$ | $\frac{1}{2}$ |
| $|0, T_1^{+1}\rangle$ | 1 | 0 | $|0, T_1^{+1}\rangle$ | 1 | 0 |
| $|0, T_1^{-1}\rangle$ | 0 | 1 | $|0, T_1^{-1}\rangle$ | 0 | 1 |

## Diagonalization of the system Hamiltonian

The system Hamiltonian $H_{\text{sys}}$ can be diagonalized separately in the sector where the molecule is in sates $|0, S_0\rangle$ or $|0, S_1\rangle$ and in other sectors. In the first sector, the Hamiltonian corre-

sponds to the quantum Rabi model. Although this model is known to has the analytical solution,[2] we numerically diagonalize the Hamiltonian in a truncated, finite-dimensional Hilbert space. The eigenstates (polaritonic states) are labeled in ascending order of their energy, where the ground and $n$-the excited polaritonic states are represented by $|0, \mathrm{GS}\rangle$ and $|0, \mathrm{P}_n\rangle$, respectively. In the second sector where the molecule is in the uncoupled states ($|0, \mathrm{T}_1\rangle$ and $|\pm 1, \mathrm{D}_0\rangle$), the eigenstates of $\hat{H}_\mathrm{sys}$, being of tensor product form, can be obtained and denoted by $|N, A; n_\mathrm{p}\rangle$, where $|N, A\rangle$ is the eigenstates of the isolated molecule and $|n_\mathrm{p}\rangle$ is the eigenstates of $\hat{H}_\mathrm{P}$ with a definite number $n_\mathrm{p}$ of plasmons ($n_\mathrm{p} = 0, 1, 2, \ldots$).

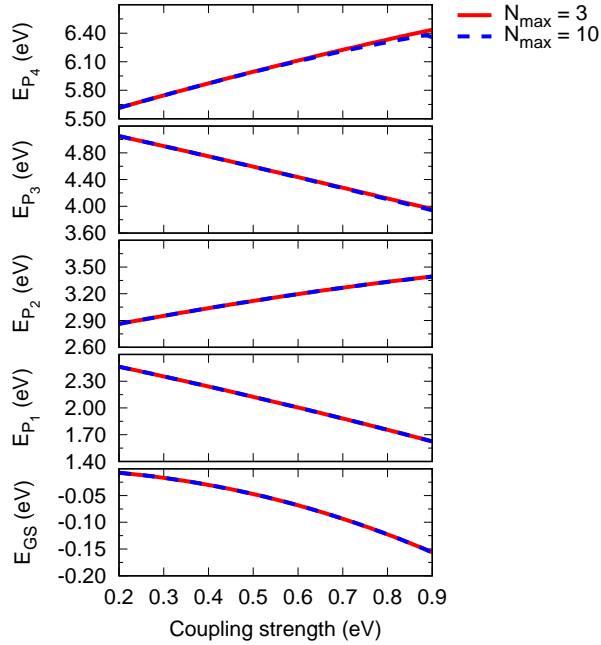

Figure S1: Energy of the polaritonic state as a function of the exciton–plasmon coupling strength $\hbar g$ for the plasmon energy $\hbar \omega_\mathrm{p} = 2.67$ eV. The energy of the first excited electronic state with the singlet spin multiplicity is set as $E_{0,\mathrm{S}_1} = 2.67$ eV. Energies of the polaritonic states $|0, \mathrm{GS}\rangle$ and $|0, \mathrm{P}_n\rangle$ are represented by $E_\mathrm{GS}$ and $E_{\mathrm{P}_n}$, respectively. Red solid and blue dashed lines show the results for the cases where the maximum number $N_\mathrm{max}$ of plasmonic quanta is set as $N_\mathrm{max} = 3$ and 10, respectively.

Figure S1 displays the energies $E_\mathrm{GS}$ and $E_{\mathrm{P}_n}$ of the ground and $n$-th excited polaritonic states as a function of the exciton–plasmon coupling strength $\hbar g$. Red solid and blue dashed lines represent the results obtained when the Hilbert space for plasmons is truncated to the

low-lying four and eleven states, respectively. These are in good agreement with each others. Thus, we truncate the Hilbert space for plasmons to the low-lying four states in numerical calculations.

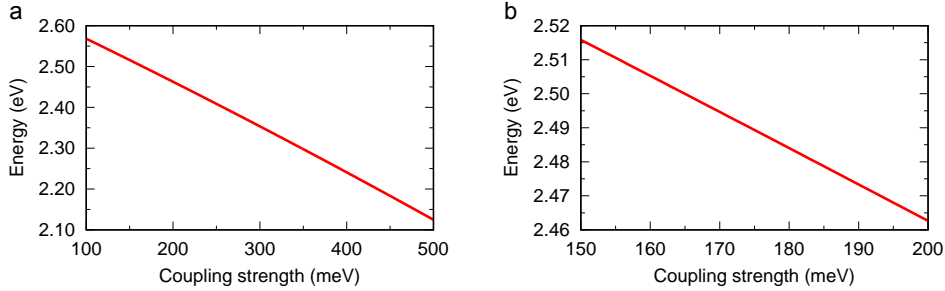

Figure S2: Energy of the first excited polaritonic state as a function of the exciton–plasmon coupling strength $\hbar g$ for the plasmon energy $\hbar\omega_p = 2.67$ eV. The exciton–plasmonc coupling strength is in the range of (a) 100 meV to 500 meV and (b) 150 meV to 200 meV. The energy of the first excited electronic state with the singlet spin multiplicity is set as $E_{0,S_1} = 2.67$ eV.

Figure S2 displays the energy $E_{P_1}$ of the first excited polaritonic states as a function of $\hbar g$ for the plasmon energy $\hbar\omega_p = 2.67$ eV. The energy of the first excited electronic state with the singlet spin multiplicity is set as $E_{0,S_1} = 2.67$ eV. In the main text, the energy of the first excited electronic state with the triplet spin multiplicity is set as $E_{0,T_1} = 2.49$ eV. Figure S2 shows that the energetic position of $E_{0,P_1}$ becomes lower than that of $E_{0,T_1}$ when $\hbar g \geq 175$ meV.

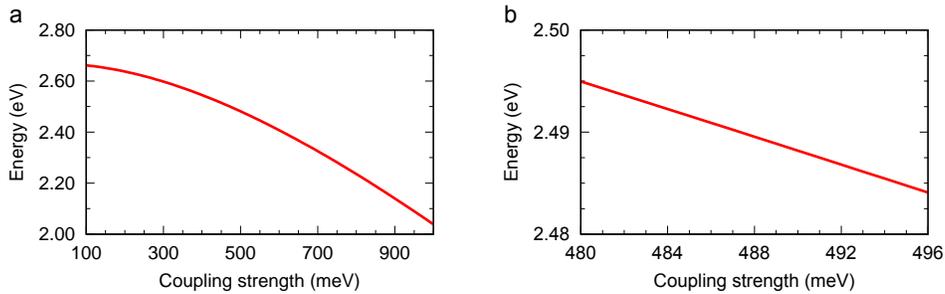

Figure S3: Energy of the first excited polaritonic state as a function of the exciton–plasmon coupling strength $\hbar g$ for the plasmon energy $\hbar\omega_p = 3.87$ eV. The exciton–plasmon coupling strength is in the range of (a) 100 meV to 1000 meV and (b) 480 meV to 496 meV. The energy of the first excited electronic state with the singlet spin multiplicity is set as $E_{0,S_1} = 2.67$ eV.

Figure S3 displays $E_{0,P_1}$ as a function of $\hbar g$ for the case of $\hbar\omega_p = 3.87$ eV. The energy

5

detuning is defined as $\Delta E = \hbar\omega_{\rm p} - E_{0,{\rm S}_1}$. In the current setting, $\Delta E$ is 1.20 eV. The energetic position of $E_{\rm P_1}$ becomes lower than that of $E_{0,{\rm T}_1}$ when $\hbar g \geq 488$ meV.

## Rate constants in the system's eigenstate representation

We express the system's eigenstates as $\{|j\rangle\}$, which comprises of $|0, {\rm GS}\rangle$, $|0, {\rm P}_n\rangle$, $|0, {\rm T}_1^m; n_{\rm p}\rangle$, and $|\pm 1, {\rm D}_0^\sigma; n_{\rm p}\rangle$. The rate constants in the eigenstate representation are expressed as

$$\Gamma^K_{j_+;j} = \sum_{N,A,B} \sum_{n_{\rm p}} |\langle j_+ | N, A; n_{\rm p}\rangle \langle N-1, B; n_{\rm p} | j\rangle|^2 \Gamma_K \nu_{N,A;N-1,B}, \quad ({\rm S13})$$

$$\Gamma^K_{j_-;j} = \sum_{N,A,B} \sum_{n_{\rm p}} |\langle j_- | N-1, B; n_{\rm p}\rangle \langle N, A; n_{\rm p} | j\rangle|^2 \Gamma_K \nu_{N,A;N-1,B}, \quad ({\rm S14})$$

where $|j_\pm\rangle$ stands for the state in which the number of electrons is one more or less than that of $|j\rangle$.

## Derivation of quantum master equation

We summarize the quantum master equation used to obtain the steady state of the system. The dynamics of the system coupled to surrounding environments are well described by the reduced density operator $\hat{\rho}(t)$, which is the partial trace of the total density operator $\hat{\rho}_{\rm tot}(t)$ over the environmental degrees of freedom, $\hat{\rho}(t) = {\rm tr}_{\rm L+R+rad+B}[\hat{\rho}_{\rm tot}(t)]$. In a weak system–environment coupling regime, it is typically relevant to employ the Born–Markov approximation.[3] The quantum master equation to describe time-evolution of the density matrices in the system's eigenstate representation, $\rho_{jk}(t) = \langle j|\hat{\rho}(t)|k\rangle$ is written as

$$\frac{d}{dt}\rho_{jk}(t) = -i\omega_{jk}\rho_{jk}(t) + \sum_{m,n}\left[R^{({\rm rad})}_{jk,mn} + \sum_{K={\rm L,R}} R^{(K)}_{jk,mn} + R^{({\rm PB})}_{jk,mn} + \sum_{\bf N} R^{({\rm MB},{\bf N})}_{jk,mn}\right]\rho_{mn}(t), \quad ({\rm S15})$$

where $\hbar\omega_{jk}$ is the energy gap between two eigenstates, $|j\rangle$ and $|k\rangle$, and the notation $\mathbf{N} = (N, A, B)$ has been introduced. The first term originates from the Hamiltonian of the system composed of the molecular DOFs and LSPs in the plasmonic nanocavity. The second, third, fourth, and fifth terms describe the relaxation dynamics induced by the plasmon–radiation field, molecule–electrode $K$, plasmon–bosonic bath, and molecule–bosonic bath coupling, respectively.

**Relaxation tensors for the bosonic baths and the radiation field.** The relaxation tensor for the bosonic baths and the radiation field is expressed in terms of correlation functions of environmental DOFs,[4]

$$\begin{aligned}R^{(\lambda)}_{jk,mn} =& + \frac{1}{\hbar^2}\tilde{C}_\lambda[\omega_{jm}]\langle j|\hat{A}_\lambda|m\rangle\langle n|\hat{A}_\lambda|k\rangle \\ &+ \frac{1}{\hbar^2}\tilde{C}^*_\lambda[\omega_{kn}]\langle j|\hat{A}_\lambda|m\rangle\langle n|\hat{A}_\lambda|k\rangle \\ &- \frac{1}{\hbar^2}\sum_l \tilde{C}_\lambda[\omega_{lm}]\delta_{nk}\langle j|\hat{A}_\lambda|l\rangle\langle l|\hat{A}_\lambda|m\rangle \\ &- \frac{1}{\hbar^2}\sum_l \tilde{C}^*_\lambda[\omega_{ln}]\delta_{jm}\langle n|\hat{A}_\lambda|l\rangle\langle l|\hat{A}_\lambda|k\rangle,\end{aligned} \quad (S16)$$

where $\lambda$ indicates the type of the environments. In Eq. (S16), $\hat{A}_\lambda$ denotes the system part of a system–environment coupling operator, i.e., $\hat{A}_{\mathrm{PB}} = \hat{A}_{\mathrm{rad}} = \hat{a}_{\mathrm{p}} + \hat{a}^\dagger_{\mathrm{p}}$ and $\hat{A}_{\mathrm{MB},\mathbf{N}} = |N, A\rangle\langle N, B| + \mathrm{h.c.}$. On the other hand, $\tilde{C}_\lambda[\omega]$ is the Fourier–Laplace transform of the correlation function of the environment $\lambda$, which is defined by $C_\lambda(t) = \langle \hat{X}_\lambda(t)\hat{X}_\lambda(0)\rangle_\lambda$ for the environmental part in a system–environment coupling operator, i.e., $\hat{X}_{\mathrm{PB}} = \sum_\beta u^{\mathrm{p}}_\beta(\hat{f}_\beta + \hat{f}^\dagger_\beta)$, $\hat{X}_{\mathrm{MB},\mathbf{N}} = \sum_\beta u^{\mathrm{M}}_{\beta,\mathbf{N}}(\hat{f}_\beta + \hat{f}^\dagger_\beta)$ and $\hat{X}_{\mathrm{rad}} = \sum_\alpha U_\alpha(\hat{b}_\alpha + \hat{b}^\dagger_\alpha)$. The bracket $\langle\ldots\rangle_\lambda$ stands for the average over the canonical density operator of the environment $\lambda$ with the inverse temperature $\beta = 1/k_{\mathrm{B}}T$. The correlation function is generally expressed in terms of the spectral density, $J_\lambda(\omega)$ to characterize the frequency dependence of the coupling constants, $u^{\mathrm{p}}_\beta$, $u^{\mathrm{M}}_{\beta,\mathbf{N}}$, and $U_\alpha$,[4] $C_\lambda(t) = (\hbar/\pi)\int_{-\infty}^\infty d\omega\, J_\lambda(\omega)[n_{\mathrm{BE}}(\omega) + 1]e^{-i\omega t}$, where $n_{\mathrm{BE}}(\omega) = 1/(e^{\beta\hbar\omega} - 1)$ is the Bose–Einstein distribution function. This yields $\mathrm{Re}\,\tilde{C}_\lambda[\omega] = \hbar J_\lambda(\omega)n_{\mathrm{BE}}(\omega)$ and

7

Im $\tilde{C}_\lambda[\omega] = (\hbar/\pi)\mathcal{P} \int_{-\infty}^{\infty} d\omega' J_\lambda(\omega') n_{\text{BE}}(\omega')/(\omega - \omega')$, where $\mathcal{P}$ stands for the principal value of the integral. The real part describes the relaxation dynamics, whereas the imaginary part corresponds to the Lamb shift. In cases of weak system-environment coupling, the Lamb shift does not dramatically alter the relaxation dynamics.[3,5] Therefore, the imaginary part is neglected in this work, and the following relation is employed in Eq. (S16):

$$\tilde{C}_\lambda[\omega] \simeq \hbar J_\lambda(\omega) n_{\text{BE}}(\omega). \tag{S17}$$

In the full form of quantum master equation, diagonal and off-diagonal elements of the reduced density matrix are generally coupled to each other, exhibiting complex quantum dynamics.[5] Through the use of the so-called secular approximation, however, the diagonal and off-diagonal elements are separated and thus the equation can be reduced to a simpler form. The approximation can be justified when the system-environment coupling is sufficiently weak so that the condition, $|R^{(\lambda)}_{jk,jk}| \ll |\omega_{jk} - \omega_{mn}|$ holds. In this work, the plasmon–bosonic bath coupling and the molecule–bosonic bath coupling are considered sufficiently small compared with the characteristic energy scale of the system. The norms of the relaxation tensors are evaluated through the values of $\gamma_{\text{m,nr}}$ and $\gamma_{\text{p,nr}}$ as the order of $10^{-4}$ eV or less. As presented in Table 1 of the main text, on the other hand, the energy gaps between the system's eigenstates are on the order of sub-eV or eV. These evaluations enable us to apply the secular approximation to $R^{(\text{PB})}_{jk,mn}$ and $R^{(\text{MB},\mathbf{N})}_{jk,mn}$. The relaxation tensor needs to be computed with a specific form of the spectral density function via Eq. (S17). Under the secular approximation, however, the tensor elements can be evaluated with the use of experimentally accessible parameters such as the nonradiative decay constants, which are expressed in terms of the spectral density. The relaxation tensor for the whole bosonic baths, $R^{(\text{B})}_{jk,mn} = R^{(\text{PB})}_{jk,mn} + \sum_{\mathbf{N}} R^{(\text{MB},\mathbf{N})}_{jk,mn}$ is thus expressed as

$$R^{(\text{B})}_{jj,mm} = \sum_{N,A,B} |\langle j|N,A\rangle\langle N,B|m\rangle|^2 \gamma_{\text{m,nr}} + |\langle j|\hat{a}_{\text{p}}|m\rangle|^2 \gamma_{\text{p,nr}}, \tag{S18}$$

where $\gamma_{\text{m,nr}}$ is the rate of the nonradiative decay from $|0, T_1\rangle$ to $|0, S_0\rangle$, and $\gamma_{\text{p,nr}}$ is the rate of the nonradiative decay of plasmons. For $j \neq k$ or $m \neq n$, the tensor elements of $R_{jk,mn}^{(\text{B})}$ are set to zero.

Contrary to the plasmon–bosonic bath and molecule–bosonic bath couplings, the plasmon–radiation field coupling is not sufficiently weak. The numerically evaluated relaxation tensor $R_{jk,mn}^{(\text{rad})}$ contains elements whose magnitudes are comparable to the characteristic energy of the system dynamics. This is consistent with Ref. 6 that demonstrated the timescale of the dynamics induced by the plasmon–radiation field coupling was on the order of 10 fs, which is comparable to the characteristic timescale of the system. Therefore, the secular approximation is not applicable to $R_{jk,mn}^{(\text{rad})}$, and thus, the full form of the relaxation tensor needs to be considered on the basis of the spectral density function, $J_{\text{rad}}(\omega)$ via Eq. (S17). Generally, the energy dependence of the plasmon–radiation field coupling has a complicated form that could markedly vary depending on materials and shapes of the plasmonic nanocavity. In this study, we consider a single representative plasmonic mode that dominantly contributes to the formation of the polaritonic states and the optical response. To investigate general characteristics of the plasmon–radiation field coupling, moreover, we neglect the spatial extent of the electric polarization of plasmons. Under the assumptions, the plasmon–radiation field coupling corresponds to the coupling of a point dipole to the radiation field, which leads to a super-Ohmic spectral density.[7] For simplicity, we employ the spectral density of the form, $J_{\text{rad}}(\omega) = \zeta \omega^3$, where $\zeta$ quantifies the strength of the plasmon–radiation field coupling and needs to be evaluated with experimentally accessible parameters. The longitudinal relaxation time for population $\rho_{kk}$ and the transverse relaxation time for coherence $\rho_{jk}$ are expressed in terms of the relaxation tensors as $T_1 = 1/\sum_{j(\neq k)}(R_{jj,kk} + R_{kk,jj})$ and $T_2 = 1/|\text{Re}\, R_{jk,jk}|$.[4] In this study, the low-lying four eigenstates of the plasmons are considered; however, it is reasonable to assume that the lowest two eigenstates dominantly contribute to the relaxation dynamics. The longitudinal and transverse relaxation times of the plasmon can be approximately expressed as $T_{1,\text{p}} = \hbar/(2\zeta\omega_{\text{p}}^3)$ and $T_{2,\text{p}} = \hbar/(\zeta\omega_{\text{p}}^3)$, which satisfy the well-

known phenomenological relation, $1/T_2 = 1/2T_1$. Consequently, the value of $\zeta$ is obtained as $\zeta = \hbar/(T_{2,\mathrm{p}}\omega_\mathrm{p}^3)$, and the spectral density that will be used to calculate the relaxation tensor is determined as $J_\mathrm{rad}(\omega) = (\hbar/T_{2,\mathrm{p}})(\omega/\omega_\mathrm{p})^3$.

**Relaxation tensors for the electrodes.** On the basis of Ref.,[8] the relaxation tensor for the electrodes is given by

$$
\begin{aligned}
R^{(K)}_{jk,mn} = \sum_{\mathbf{N}} &\left[ \langle j|(\hat{\alpha}^K_{\mathbf{N}})^\dagger|m\rangle\langle n|N-1,B\rangle\langle N,A|k\rangle + \langle j|N,A\rangle\langle N-1,B|m\rangle\langle n|\hat{\alpha}^K_{\mathbf{N}}|k\rangle \right.\\
&\left. + \langle j|\hat{\beta}^K_{\mathbf{N}}|m\rangle\langle n|N,A\rangle\langle N-1,B|k\rangle + \langle j|N-1,B\rangle\langle N,A|m\rangle\langle n|(\hat{\beta}^K_{\mathbf{N}})^\dagger|k\rangle \right]\\
- \delta_{nk}\sum_{\mathbf{N}}\sum_l &\left[ \langle j|N-1,B\rangle\langle N,A|l\rangle\langle l|(\hat{\alpha}^K_{\mathbf{N}})^\dagger|m\rangle + \langle j|N,A\rangle\langle N-1,B|l\rangle\langle l|\hat{\beta}^K_{\mathbf{N}}|m\rangle \right]\\
- \delta_{jm}\sum_{\mathbf{N}}\sum_l &\left[ \langle n|\hat{\alpha}^K_{\mathbf{N}}|l\rangle\langle l|N,A\rangle\langle N-1,B|k\rangle + \langle n|(\hat{\beta}^K_{\mathbf{N}})^\dagger|l\rangle\langle l|N-1,B\rangle\langle N,A|k\rangle \right], \quad \text{(S19)}
\end{aligned}
$$

with

$$
\hat{\alpha}^K_{\mathbf{N}} = \sum_{\mathbf{N}'}\sum_{j,k}\frac{1}{2}\Gamma^K_{\mathbf{NN}'}(\varepsilon_{kj})f_K(\varepsilon_{kj})\langle j|N'-1,B'\rangle\langle N',A'|k\rangle|j\rangle\langle k|, \quad \text{(S20)}
$$

$$
\hat{\beta}^K_{\mathbf{N}} = \sum_{\mathbf{N}'}\sum_{j,k}\frac{1}{2}\Gamma^K_{\mathbf{NN}'}(\varepsilon_{kj})[1-f_K(\varepsilon_{kj})]\langle j|N'-1,B'\rangle\langle N',A'|k\rangle|j\rangle\langle k|, \quad \text{(S21)}
$$

where $f_K(\varepsilon) = 1/[e^{\beta(\varepsilon-\mu_K)}+1]$ is the Fermi distribution function with chemical potential $\mu_K$, and $\Gamma^K_{\mathbf{NN}'}(\varepsilon) = (2\pi/\hbar)\sum_{k\in K}\sum_\sigma V_{Kk\sigma,\mathbf{N}}V^*_{Kk\sigma,\mathbf{N}'}\delta(\varepsilon-\epsilon_{Kk\sigma})$ has been introduced. The magnitude of $\Gamma^K_{\mathbf{NN}'}(\varepsilon)$ is reasonably assumed to be equal to or less than that of $\Gamma^K_{\mathbf{NN}}(\varepsilon)$. Furthermore, the magnitude of $\Gamma^K_{\mathbf{NN}}(\varepsilon)$ is equal to or less than the value of $\Gamma_K$, as presented in SM. Hence, the norm of the relaxation tensor is evaluated through the values of $\Gamma_K$ as the order of $10^{-4}$ eV, which leads to the applicability of the secular approximation. In the same

manner as in Eq. (S18), the necessary relaxation tensor elements can be simply expressed as

$$R^{(K)}_{j_+j_+,jj} = \Gamma^K_{j_+;j} f_K(\varepsilon_{j_+} - \varepsilon_j), \tag{S22}$$

$$R^{(K)}_{j_-j_-,jj} = \Gamma^K_{j_-;j} [1 - f_K(\varepsilon_j - \varepsilon_{j_-})], \tag{S23}$$

where the definition of $\Gamma^K_{j;m}$ is given in SM. Note that $|j_\pm\rangle$ stands for the system's eigenstate in which the number of electrons is one more or less than that of $|j\rangle$. For $j \neq k$ or $m \neq n$, the tensor elements of $R^{(K)}_{jk,mn}$ are set to zero.

## Formulation of photon flux

An average population of photons in the radiation field is evaluated as $\langle N(t)\rangle = \text{tr}[\sum_\alpha \hat{b}^\dagger_\alpha \hat{b}_\alpha \hat{\rho}_{\text{tot}}(t)]$, and therefore, the rate of change in the population, $J_{\text{ph}}(t) = (d/dt)\langle N(t)\rangle$ is obtained as

$$J_{\text{ph}}(t) = -\frac{i}{\hbar} \sum_\alpha U_\alpha \text{tr}[\tilde{A}_{\text{rad}}(t)\tilde{b}^\dagger_\alpha(t)\tilde{\rho}_{\text{tot}}(t)] + \text{c.c.}. \tag{S24}$$

To derive an analytical expression of Eq. (S24), a source term $f(t)$ is added to the plasmon–radiation field interaction Hamiltonian and the expectation value of an operator that includes $\hat{b}_\alpha$ and $\hat{b}^\dagger_\alpha$ is obtained through the functional derivative of the density operator with respect to $f(t)$. The plasmon–radiation field interaction Hamiltonian with the source term introduced as

$$\hat{V}_{\text{P,rad},f} = [\hat{A}_{\text{rad}} + f(t)]\sum_\alpha U_\alpha \hat{b}^\dagger_\alpha + \hat{A}^\dagger_{\text{rad}} \sum_\alpha U_\alpha \hat{b}_\alpha. \tag{S25}$$

We postulate that the source term affects only the ket side of the density operator, and we let $\hat{\rho}_{\text{tot},f}(t)$ denote the total density operator with the source term. The functional derivative of $\hat{\rho}_{\text{tot},f}(t)$ with respect to $f(t)$ yields $[\delta/\delta f(t)]\tilde{\rho}_{\text{tot},f}(t)|_{f=0} = -(i/\hbar)\sum_\alpha U_\alpha \tilde{b}^\dagger_\alpha(t)\tilde{\rho}_{\text{tot}}(t)$, where $\tilde{O}(t)$ indicates the interaction representation of an arbitrary operator $\hat{O}$ with respect to

$\hat{H}_{\text{sys}} + \hat{H}_{\text{env}}$. As the consequence, Eq. (S24) is expressed as

$$J_{\text{ph}}(t) = \text{tr}_{\text{sys}} \left[ \tilde{A}_{\text{rad}}(t) \frac{\delta}{\delta f(t)} \tilde{\rho}_f(t) \bigg|_{f=0} \right] + \text{c.c.}, \quad (S26)$$

where $\tilde{\rho}_f(t) = \text{tr}_{\text{L+R+rad+B}}[\tilde{\rho}_{\text{tot},f}(t)]$ is the reduced density operator with the source term. We assume that the time evolution of the subsystem can be relevantly described under the second-order perturbative approximation with respect to $\hat{V}$ and the Markovian approximation, leading to

$$J_{\text{ph}}(t) \simeq \int_0^\infty d\omega \frac{J_{\text{rad}}(\omega)}{\pi \hbar} \int_0^\infty d\tau \, e^{i\omega\tau} \Big\{ [n_{\text{BE}}(\omega) + 1] \text{tr}_{\text{sys}}[\tilde{A}_{\text{rad}}(t) \tilde{\rho}(t) \tilde{A}_{\text{rad}}^\dagger(t-\tau)]$$
$$- n_{\text{BE}}(\omega) \text{tr}_{\text{sys}}[\tilde{A}_{\text{rad}}(t) \tilde{A}_{\text{rad}}^\dagger(t-\tau) \tilde{\rho}(t)] \Big\} + \text{c.c.}. \quad (S27)$$

In the eigenstate representation of the system, Eq. (S27) is recast into

$$J_{\text{ph}}(t) \simeq \frac{2}{\hbar} \sum_{l,m,n;\omega_{nl}>0} J_{\text{rad}}(\omega_{nl}) \text{Re} \Big\{ [n_{\text{BE}}(\omega_{nl}) + 1] \langle l|\hat{A}_{\text{rad}}|m\rangle \langle m|\hat{\rho}(t)|n\rangle \langle n|\hat{A}_{\text{rad}}^\dagger|l\rangle$$
$$- n_{\text{BE}}(\omega_{nl}) \langle m|\hat{A}_{\text{rad}}|n\rangle \langle n|\hat{A}_{\text{rad}}^\dagger|l\rangle \langle l|\hat{\rho}(t)|m\rangle \Big\}, \quad (S28)$$

where the Lamb shift in the electronic energies are neglected in the same fashion as in Eq. (S16).

# Derivation of analytic expression of the electroluminescence efficiency

We derive an analytic expression of the electroluminescence efficiency $\eta$ by solving a rate equation for the populations of the respective eigenstates of the system. The system's eigenstates are given by $|0, \text{GS}\rangle$, $|0, \text{P}_n\rangle$, $|0, \text{T}_1^{m=0,\pm 1}; n_{\text{p}}\rangle$, $|+1, \text{D}_0^{\sigma=\pm 1/2}; n_{\text{p}}\rangle$, $|-1, \text{D}_0^{\sigma=\pm 1/2}; n_{\text{p}}\rangle$ with $n = 1, 2, \ldots$ and $n_{\text{p}} = 0, 1, 2, \ldots$. The transition from $|N, i\rangle$ to $|N-1, j\rangle$ accompanies an

electron transfer between the molecule and electrode $K$, and the transition rate is given by $k^K_{N-1,j\leftarrow N,i} = \Gamma^K_{N-1,j;N,i}[1 - f_K(\epsilon_{N,i} - \epsilon_{N-1,j})]$, where $\epsilon_{N,i}$ is the eigenenergy of $|N,i\rangle$ and $\Gamma^K_{N-1,j;N,i}$ is defined in SM. Similarly, the transition rate from $|N,i\rangle$ to $|N+1,j\rangle$ is written as $k^K_{N+1,j\leftarrow N,i} = \Gamma^K_{N+1,j;N,i} f_K(\epsilon_{N+1,j} - \epsilon_{N,i})$. The transition rate from $|0,P_1\rangle$ to $|0,GS\rangle$ is written by $k_{0,GS\leftarrow 0,P_1} = \kappa k_{p,r} + \kappa k_{p,nr}$, where $\kappa k_{p,r}$ and $\kappa k_{p,nr}$ represent the radiative and nonradiative decay rates, respectively. As shown in Table 1 of the main text, we assume that the nonradiative decay rate $\hbar\gamma_{m,nr}$ of the molecule is much smaller than that $\hbar\gamma_{p,nr}$ of plasmons. Therefore, in deriving $k_{0,GS\leftarrow 0,P_1}$, we neglect the effect of $\hbar\gamma_{m,nr}$ on $k_{0,GS\leftarrow 0,P_1}$.

The numerical results in the main text demonstrated that the selective excitation of $|0,P_1\rangle$ without populating $|0,T_1;0\rangle$ leads to efficient electroluminescence. To analyse the results, we consider a case in which the following two conditions are satisfied: the coupling strength $\hbar g$ is strong enough to realise $E_{P_1} \leq E_{0,T_1}$, and the bias voltage is in the range of $V_{P_1} < V < V_{T_1}$. The rate equations for this case are written as

$$\frac{d}{dt}P_{0,GS}(t) = -\sum_{K=L,R}\sum_{\sigma=\pm 1/2}\sum_{N=\pm 1} k^K_{N,D_0^\sigma;0\leftarrow 0,GS}P_{0,GS}(t) + k_{0,GS\leftarrow 0,P_1}P_{0,P_1}(t)$$
$$+ \sum_{K=L,R}\sum_{\sigma=\pm 1/2}\sum_{N=\pm 1} k^K_{0,GS\leftarrow N,D_0^\sigma;0}P_{N,D_0^\sigma;0}(t), \quad (S29)$$

$$\frac{d}{dt}P_{0,P_1}(t) = -\left(k_{0,GS\leftarrow 0,P_1} + \sum_{K=L,R}\sum_{\sigma=\pm 1/2}\sum_{N=\pm 1} k^K_{N,D_0^\sigma;0\leftarrow 0,P_1}\right)P_{0,P_1}(t)$$
$$+ \sum_{K=L,R}\sum_{\sigma=\pm 1/2}\sum_{N=\pm 1} k^K_{0,P_1\leftarrow N,D_0^\sigma;0}P_{N,D_0^\sigma;0}(t), \quad (S30)$$

$$\frac{d}{dt}P_{N,D_0^\sigma;0}(t) = \sum_{K=L,R}\left[k^K_{N,D_0^\sigma;0\leftarrow 0,GS}P_{0,GS}(t) + k^K_{N,D_0^\sigma;0\leftarrow 0,P_1;0}P_{0,P_1}(t)\right.$$
$$\left. - \left(k^K_{0,GS\leftarrow N,D_0^\sigma;0} + k^K_{0,P_1;0\leftarrow N,D_0^\sigma;0}\right)P_{N,D_0^\sigma;0}(t)\right], \quad (S31)$$

where $P_{0,GS}(t)$, $P_{0,P_1}(t)$, and $P_{N,D_0^\sigma;0}(t)$ are the populations of $|0,GS\rangle$, $|0,P_n\rangle$, and $|N,D_0^\sigma;0\rangle$, respectively. In the equations, we neglect the transitions from/to $|0,P_n\rangle$, $|0,T_1^m;0\rangle$, $|0,T_1^m;n_p\rangle$, and $|N,D_0^\sigma;n_p\rangle$ with $n \geq 2$ and $n_p = 1,2,\ldots$, because the rates of these transitions are van-

13


\end{document}

%% file: main.bbl
\providecommand{\latin}[1]{#1}
\makeatletter
\providecommand{\doi}
  {\begingroup\let\do\@makeother\dospecials
  \catcode`\{=1 \catcode`\}=2 \doi@aux}
\providecommand{\doi@aux}[1]{\endgroup\texttt{#1}}
\makeatother
\providecommand*\mcitethebibliography{\thebibliography}
\csname @ifundefined\endcsname{endmcitethebibliography}
  {\let\endmcitethebibliography\endthebibliography}{}